%% file: main.tex
\newif\ifdraft

\newif\ifpreprint
\preprinttrue

\documentclass[conference]{IEEEtran}

\usepackage{booktabs}
\usepackage{amsmath,amssymb,amsfonts}
\usepackage{algorithmic}
\usepackage{graphicx}
\usepackage{textcomp}
\usepackage{xcolor}
\usepackage[hidelinks]{hyperref}
\usepackage{listings}
\usepackage{siunitx}
\usepackage{multirow}
\usepackage{xspace}
\usepackage{balance}
\usepackage[numbers,compress]{natbib}



\ifpreprint
    \usepackage[T1]{fontenc}
    \usepackage{inconsolata}
\fi

\definecolor{backgroundColour}{HTML}{FAFAFA}
\definecolor{keywordclr}{HTML}{3F51B5}
\definecolor{commentclr}{HTML}{757575}
\definecolor{stringsclr}{HTML}{279049}
\definecolor{fnctionclr}{rgb}{0.467, 0, 0.533}
\definecolor{builtinclr}{rgb}{0.35, 0, 0.533}
\definecolor{symbolsclr}{rgb}{0.5, 0.25, 0.25}   
\definecolor{numbersclr}{rgb}{0.8, 0.2, 0}
\definecolor{bckgrndclr}{rgb}{0.91, 0.95, 0.95}
\lstdefinestyle{PythonStyle}{
    language=Python,
    backgroundcolor=\color{backgroundColour},
    keywordstyle=\color{keywordclr}\bfseries,
    stringstyle=\color{stringsclr},
    commentstyle=\color{commentclr}\itshape,
    upquote=true,
    basicstyle=\ttfamily\linespread{0.9}\scriptsize,
    breakatwhitespace=false,
    breaklines=true,
    captionpos=b,
    keepspaces=true,
    numbers=left,
    numbersep=5pt,
    numberstyle=\color{commentclr}\ttfamily\tiny,
    showspaces=false,
    showstringspaces=false,
    showtabs=false,
    tabsize=2,
    xleftmargin=1.25em,
    frame=single,
    framexleftmargin=1.25em,
    morekeywords={assert,with,as}
}
\ifdraft
  \newcommand{\todo}[1]{{\textcolor{red}{ TODO: #1 }}}
  \newcommand{\ian}[1]{{\textcolor{green}{ Ian: #1 }}}
  \newcommand{\greg}[1]{{\textcolor{purple}{ Greg: #1 }}}
  \newcommand{\val}[1]{{\textcolor{orange}{ Valerie: #1 }}}
  \newcommand{\kyle}[1]{{\textcolor{red}{ Kyle: #1 }}}
  \newcommand{\haochen}[1]{{\textcolor{cyan}{ Haochen: #1 }}}
\else
  \newcommand{\todo}[1]{}
  \newcommand{\ian}[1]{}
  \newcommand{\greg}[1]{}
  \newcommand{\val}[1]{}
  \newcommand{\kyle}[1]{}
  \newcommand{\haochen}[1]{}
\fi

\newcommand{\name}{\textsc{TaPS}\xspace}

\begin{document}

\title{\name{}: A Performance Evaluation Suite for Task-based Execution Frameworks}

\author{
\IEEEauthorblockN{
J. Gregory Pauloski,\IEEEauthorrefmark{1}\IEEEauthorrefmark{2}
Valerie Hayot-Sasson,\IEEEauthorrefmark{1}\IEEEauthorrefmark{2}
Maxime Gonthier,\IEEEauthorrefmark{1}\IEEEauthorrefmark{2}
Nathaniel Hudson,\IEEEauthorrefmark{1}\IEEEauthorrefmark{2}\\
Haochen Pan,\IEEEauthorrefmark{1}
Sicheng Zhou,\IEEEauthorrefmark{1}
Ian Foster,\IEEEauthorrefmark{1}\IEEEauthorrefmark{2}
and Kyle Chard\IEEEauthorrefmark{1}\IEEEauthorrefmark{2}
}
\IEEEauthorblockA{
    \IEEEauthorrefmark{1}Department of Computer Science, University of Chicago, Chicago, IL, USA}
\IEEEauthorblockA{
    \IEEEauthorrefmark{2}Data Science and Learning Division, Argonne National Laboratory, Lemont, IL, USA
}
}

\maketitle

\begin{abstract}
Task-based execution frameworks, such as parallel programming libraries, computational workflow systems, and function-as-a-service platforms, enable the composition of distinct tasks into a single, unified application designed to achieve a computational goal.
Task-based execution frameworks abstract the parallel execution of an application's tasks on arbitrary hardware.
Research into these task executors has accelerated as computational sciences increasingly need to take advantage of parallel compute and/or heterogeneous hardware.
However, the lack of evaluation standards makes it challenging to compare and contrast novel systems against existing implementations.
Here, we introduce \name{}, the Task Performance Suite, to support continued research in parallel task executor frameworks.
\name{} provides (1) a unified, modular interface for writing and evaluating applications using arbitrary execution frameworks and data management systems and (2) an initial set of reference synthetic and real-world science applications.
We discuss how the design of \name{} supports the reliable evaluation of frameworks and demonstrate \name{} through a survey of benchmarks using the provided reference applications.
\end{abstract}

\begin{IEEEkeywords}
Performance Evaluation, Parallel Computing, Open-Source, Python, Workflows
\end{IEEEkeywords}

\section{Introduction}
\label{sec:introduction}

Task-based execution frameworks, such as Dask~\cite{rocklin2015dask}, Parsl~\cite{babuji19parsl}, and Ray~\cite{moritz2018ray}, have enabled many advances across the sciences.
These \emph{task executors} manage the complexities of executing the tasks comprising an application in parallel across arbitrary hardware.
Decoupling the application logic (e.g., what tasks to perform, how data flow between tasks) from the execution details (e.g., scheduling systems or communication protocols) simplifies development and results in applications which are portable across diverse systems.
Task executors come in many forms, from a simple pool of processes to sophisticated workflow management systems (WMSs), and the rapid increase in the use of task-based applications across the computational sciences has spurred further research in the area.

Consistent and reliable benchmarking is fundamental to evaluating advances within a field over time.
Benchmarks and other performance evaluation systems offer a common ground and objective metrics that enable researchers to assess the efficiency, performance, scalability, and robustness of their solutions under controlled conditions.
Benchmarks foster transparency and reproducibility, ensuring that results can be consistently replicated and verified by others in the field.
This, in turn, accelerates the pace of innovation as researchers can identify best practices, optimize existing methods, and uncover new areas for improvement.
Benchmarks facilitate meaningful comparisons between competing approaches---a valuable aspect for researchers, reviewers, and readers alike.

Access to open source benchmarks democratizes research, and many fields have found great success through the creation of standards.
LINPACK~\cite{dongarra1988linpack}, for example, is used to evaluate the floating point performance of hardware systems.
The Transaction Processing Performance Council (TPC)~\cite{tpc} provides a variety of standard benchmarks for database systems, and UnixBench~\cite{unixbench} can evaluate basic performance of Unix-like systems from file copies to system call overheads.
Machine learning (ML) has demonstrated this success with benchmarks for every level of the ML stack.
MLPerf~\cite{mattson2020mlperf,reddi2020mlperf} has continued to support the development of ML hardware and frameworks.
Novel algorithms are compared against prior work by using open source datasets, as exemplified by the Papers with Code Leaderboards~\cite{paperswithcode} that comprise results of tens of thousands of papers across thousands of datasets.

However, the parallel application and workflows communities lack such established benchmarks.
The NAS parallel benchmarks date back to the 1990s~\cite{bailey1995parallel}.
For workflows, with the exception of a few common applications (e.g., Montage~\cite{montage,deelman2008montageworkflow}), papers typically evaluate their solutions on purpose-built synthetic benchmarks or forks of real world science applications.
Unfortunately, the ad hoc nature of these solutions means that the code is often not open sourced, not maintained beyond publication of the corresponding paper, or so specific to an implementation that it is challenging to appropriately compare against in later works.
Recent work has introduced a standard for recording execution traces and tools for analyzing those traces~\cite{colman2022wfcommons}, but there remains a need for realistic reference applications for benchmarking.

To address these challenges, we introduce \name{}, the Task Performance Suite, a standardized framework for evaluating task-based execution frameworks against synthetic and realistic science applications.
With \name{}, applications can be written in a framework-agnostic manner and then benchmarked using any one of many supported task executors and data management systems.
We make the following contributions:
\begin{enumerate}
    \item \name{}, a standardized benchmarking framework for task-based applications with an extensible plugin system for comparing task executors and data management systems.
    \name{} is available at \url{https://github.com/proxystore/taps}.
    \item Support for popular task executors (Dask, Globus Compute, Parsl, Ray, and TaskVine) and data management systems (shraed file systems and ProxyStore).
    \item Reference implementations within \name{} for six real (Cholesky factorization, protein docking, federated learning, MapReduce, molecular design, and Montage) and two synthetic applications.
    \item Insights into the performance of the reference implementations across the supported frameworks.
\end{enumerate}

The rest of this paper is as follows:
\autoref{sec:related} discusses related work;
\autoref{sec:design} describes the design and implementation details of the \name{} framework;
\autoref{sec:applications} introduces the initial set of applications provided by \name{};
\autoref{sec:evaluation} presents our experiences using \name{} to evaluate system components; and
\autoref{sec:conclusion} summarizes our contributions and future directions.

\section{Background and Related Work}
\label{sec:related}

\emph{Task executors}, which manage the execution of tasks in parallel across distributed resources, come in many forms.
A \emph{task} refers to discrete unit of work, and tasks are combined into a larger \emph{application}.
Tasks can take data as input, produce output data, and may have dependencies with other tasks;
i.e., a dependent task cannot start until a preceding tasks completes.
Dask Distributed, Python's \texttt{ProcessPoolExecutor}, Globus Compute~\cite{chard20funcx}, Radical Pilot~\cite{alsaadi2021radical}, and Ray all provide mechanisms for executing tasks in parallel across distributed systems.

\emph{Workflow management systems} (WMSs), a subset of task executors, are designed to define, manage, and execute workflows represented by a directed acyclic graph (DAG) of tasks.
WMSs commonly provide mechanisms for automating and optimizing task flow, monitoring, and resource management.
WMSs can be categorized as supporting explicit or implicit dataflow patterns.
Explicit systems, such as Apache Airflow~\cite{airflow}, Fireworks~\cite{jain2015fireworks}, Makeflow~\cite{albrecht2012makeflow}, Nextflow~\cite{di2017nextflow}, Pegasus~\cite{deelman15pegasus}, and Swift~\cite{wilde11swift}, rely on configuration files or domain specific languages (DSLs) to statically define a DAG before execution.
Implicit systems, such as Dask Delayed, Parsl, Swift/T~\cite{wozniak13swiftt}, and TaskVine~\cite{slydelgado2023taskvine}, derive the application's dataflow through the dynamic evaluation of a procedural script.

\emph{Performance evaluation} of task executors is challenging due to a lack of standards.
Frameworks provide examples designed to aid in learning the framework, but these are often too trivial to be used in benchmarking.
Pegasus provides a catalogue of real, end-to-end scientific workflows in AI, astronomy, and bio-informatics which are suitable for benchmarking~\cite{pegasus-examples}; Dask maintains a repository of performance benchmarks~\cite{dask-benchmarks}; WorkflowHub provides a service for sharing scientific workflows~\cite{workflowhub}; and Workbench~\cite{albrecht2012makeflow}, designed for analyzing workflow patterns, was released alongside Makeflow. 
However, these reference applications and benchmarks are typically valid only for evaluating optimizations within the framework they were implemented in.
In other words, the majority of these code bases are not suitable for comparing different task executors.
This also means available benchmarks are susceptible to code rot if maintenance of the associated framework ceases.

Porting benchmark applications between frameworks is onerous when the structure and syntax is completely different.
Subtle errors in the ported implementation can lead to inaccurate comparisons between systems.
Access to datasets or sufficient compute resources for certain applications can further hinder the creation of realistic benchmarking applications.
To assuage these challenges within the workflows community, prior work~\cite{silva2014community} published a gallery of execution traces from real workloads using Pegasus, a synthetic workflow generator, and a simulator framework.
WfCommons~\cite{colman2022wfcommons} 
introduces a standardized format for representing execution logs (WfFormat), an open source package for analyzing logs and generating synthetic logs (WfGen), and a workflow execution simulator (WfSim).
WfCommons currently provides 180 execution instances from three workflow systems (Makeflow, Nextflow, and Pegasus).
Similarly, WRENCH~\cite{casanova2020wrench} provides a WMS simulation framework built on SimGrid~\cite{casanova2014simgrid}.
In contrast, an Application Skeleton supports the design and development of systems by mimicking the performance of a real application~\cite{katz2016skeleton}.

FunctionBench~\cite{jeongchul2019functionbench}, FaaSDom~\cite{pascal2020faasdom}, and SeBS~\cite{copik2021sebs} address a similar set of challenges as \name{} but in the context of cloud-hosted function-as-a-service (FaaS) platforms.
SeBS provides a benchmark specification, a general model of FaaS platforms, and an implementation of the framework and benchmarks.
This model is valuable because each benchmark is platform agnostic, relying only on the abstract FaaS model provided by SeBS.
Implementing the concrete model for a new platform need only be performed once, and then any benchmark can be executed on that platform.
Part of SeBS's platform model is support for persistent and ephemeral cloud storage systems.
Supporting the evaluation of the compute and data aspects of task-based applications is crucial, but currently lacking outside of specific areas (i.e., SeBS for FaaS).

\section{Design and Implementation}
\label{sec:design}

\name{} is a Python package that provides a common framework for writing task-based, distributed applications; a plugin system for running applications with arbitrary task executors and data management systems; and a benchmarking framework for running experiments in a reproducible manner.
We choose Python for its pervasiveness in task-based, distributed applications, and we describe here the high level concepts that make the framework possible and the implementation details.
Our goal is to create an easy-to-use framework for researchers to benchmark novel systems and an extensible framework so future applications and plugins can be incorporated into \name{}.

\subsection{Application Model}

\name{} provides a framework for the creation and execution of application benchmarks.
As described in \autoref{sec:related},
applications are composed of tasks which are the remote execution of a function which takes in some data and produces some data.
Tasks can have dependencies such that the result of one task is consumed by one or more tasks.

Supporting applications written using the explicit and implicit workflow models described in \autoref{sec:related} is challenging because the two philosophies are fundamentally at odds with each other and,
within the scope of explicit systems, the different configuration formats and use of DSLs further complicates the design of a unified, abstract task executor interface.

\name{} supports writing applications as Python code using implicit dataflow dependencies.
(Though, it is not a requirement that tasks have dataflow dependencies within an application.)
We take this approach for two reasons.
First, the scope of applications compatible with implicit models is a super-set of those compatible with explicit models.
Specifically, WFMs which use a static graph for execution are not expressive enough for writing more dynamic and procedural applications, whereas the implicit model enables arbitrarily complex applications composed through a procedural program.
Second, WMFs which use DSLs require the application design to be tightly coupled to the WMF.
This inherently makes it challenging to construct an application that is compatible with a multitude of frameworks.

\subsection{Writing Applications}

\input{figures/framework-stack}
\input{listings/application}

In \name{}, an application is composed of two parts: an \texttt{AppConfig} and an \texttt{App} class (see \autoref{fig:framework-stack}).
The \texttt{AppConfig} contains all configuration options required to execute the corresponding applications (e.g., hyperparameters, paths to datasets, or flags).
\texttt{AppConfig} exposes a \texttt{get\_app()} method which initializes an \texttt{App} instance from the user-specified configuration.
\texttt{App.run()} is the entry point to the application code and is invoked with two arguments: an \texttt{Engine} instance (discussed in detail in \autoref{sec:design:engine}) and the path to a unique directory for the current application invocation.
The \texttt{run()} method can contain arbitrary code, provided application tasks are executed via the provided \texttt{Engine} interface.

\name{} provides a CLI framework for executing application benchmarks.
For example, the \texttt{foo} application in \autoref{lst:application} is started with: \texttt{python -m \MakeLowercase{\name{}}.run --app foo \{args\}}.
An \texttt{AppConfig} can be registered with the CLI using the \texttt{@register(`app')} decorator.
This will automatically add the application's name as one of the CLI choices and add CLI arguments based on the \texttt{AppConfig} attributes.

When invoked, the CLI (1) constructs an \texttt{AppConfig} instance from the user's arguments, validating that options can be parsed into the correct type and that all required arguments are present; (2) initializes the \texttt{App} using \texttt{get\_app()}; (3) constructs an \texttt{Engine} according to user-supplied arguments; and (4) invokes \texttt{App.run()} to execute the application benchmark.
The framework automatically writes a configuration file, log files, and task record files to the run directory.
The configuration file contains a record of all configuration options used to execute the application. A configuration file path can be provided to the CLI as an alternative to CLI arguments; thus, configuration files can be shared for reproducibility. 

\subsection{Application Execution}
\label{sec:design:engine}

The \texttt{Engine} is the unified interface used by applications to execute tasks and exposes an interface similar to Python's \texttt{concurrent.futures.Executor}.
The \texttt{Engine} interface must be expressive enough to build arbitrary applications yet simple enough to incorporate third-party task executors and other plugins.
We chose to adopt a model similar to Python's \texttt{Executor} because it is a \emph{de facto} standard for managing asynchronous task execution across the Python ecosystem and many third-party libraries provide \texttt{Executor}-like implementations, including Dask Distributed, Globus Compute, Loky, TaskVine, and Parsl.
An additional benefit of this choice is that it is trivial to port applications already using an \texttt{Executor} interface into a \name{} application.

\texttt{Executor} is an abstract class with two primary methods, \texttt{submit()} and \texttt{map()}, designed to execute functions asynchronously.
The \texttt{submit()} method takes a callable object and associated arguments, schedules the callable for execution, and returns back to the client a \texttt{Future} that will eventually contain the result of the callable.
\texttt{Engine} implements both of these methods, but returns \texttt{TaskFuture} objects rather than \texttt{Future} instances.
Functionally, \texttt{TaskFuture} behaves like \texttt{Future} but includes additional functionality for performance monitoring and task dependency management.

An \texttt{Engine} is created from four components: \texttt{Executor}, \texttt{Transformer}, \texttt{Filter}, and \texttt{RecordLogger}.
This conceptual hierarchy of components in \name{} is illustrated in \autoref{fig:framework-stack}.
The dependency model approach used by the \texttt{Engine} means that component plugins can be easily created and/or swapped to compare, for example, different task executors or data management systems.
Further, the \texttt{Engine} can be extended with additional components in the future to enhance benchmarking capabilities.

\subsection{Task Executor Model}
\label{sec:design:task-executor}

The fundamental component of the \texttt{Engine} is an  \texttt{Executor}, an interface to the underlying task executor.
We choose the \texttt{Executor} model again for the same reasons as with the \texttt{Engine}.
In \autoref{sec:design:supported-task-executors}, we describe the details of each executor currently supported in \name{}. 
Similar to the \texttt{App} model, \name{} has a notion of a \texttt{ExecutorConfig}
which can be registered with the framework to automatically add argument parser groups for the specific executor.
\texttt{ExecutorConfig} has a method, \texttt{get\_executor()}, which will initialize an instance of the executor from the user specified configuration.

A limitation of Python's \texttt{Executor} interface is the lack of support for dataflow dependencies between tasks.
Some \texttt{Executor} implementations (Dask Distributed, Parsl, and TaskVine) do support implicit dataflow dependencies by passing the future of one task as input to one or more tasks, but many others (e.g., Python's \texttt{ProcessPoolExecutor} and Globus Compute) do not.
The \texttt{Engine} requires it's \texttt{Executor} to support implicit dataflow patterns with futures, so \name{} provides a \texttt{FutureDependencyExecutor} wrapper to add this functionality if needed.
This wrapper scans task inputs for futures and will delay submission of a task until the results of all input futures are available (in an asynchronous, non-blocking manner).

\subsection{Supported Task Executors}
\label{sec:design:supported-task-executors}

\input{tables/engines}

Here, we briefly describe the task executors currently supported by \name{} (summarized in 
\autoref{tab:engines}).
As previously mentioned, the plugin system makes it easy to support more executors in the future, but our initial goal is to support a broad range.
We support Python's \texttt{ProcessPoolExecutor} which provides a good baseline for low-overhead, single-node execution.
We also support the \texttt{ThreadPoolExecutor}, but this is primarily intended to support development and quick testing because Python's Global Interpreter Lock prevents true parallelism with threading.

Dask Distributed~\cite{rocklin2015dask} provides dynamic task scheduling and management across worker processes distributed across cores within a node or across several nodes.
Tasks in Dask are Python functions which operate on Python objects; the scheduler tracks these task in a dynamic DAG.
Globus Compute~\cite{chard20funcx} is a cloud-managed function-as-a-service (FaaS) platform which can execute Python functions across federated compute systems.
Globus Compute provides an \texttt{Executor} interface but does not manage dependencies between functions.
Parsl~\cite{babuji19parsl} is a parallel programming library for Python with comprehensive dataflow management capabilities.
Parsl supports many execution models including local compute, remote compute, and batch scheduling systems.
Ray~\cite{moritz2018ray} is a general purpose framework for executing task-parallel and actor-based computations on distributed systems in a scalable and fault tolerant manner.
TaskVine~\cite{slydelgado2023taskvine} executes dynamic DAG workflows with a focus on data management features including transformation, distribution, and task data locality.

\subsection{Task Data Model}

Optimizing the transfer of task data and placement of tasks according to where data reside is a core feature of many task executors.
To support further research into data management, \name{} supports a plugin system for \emph{data transformers}.
A data transformer is an object that implements the \texttt{Transformer} protocol.
This protocol defines two methods: \texttt{transform} which takes an object and returns an identifier, and \texttt{resolve}, the inverse of \texttt{transform}, which takes an identifier and returns the corresponding object.
Data transformer implementations can implement object identifiers in any manner, provided identifier instances are serializable.
For example, an identifier could simply be a UUID corresponding to a database entry containing the serialized object.

A \texttt{Filter} is a callable object, e.g., a function, that takes an object as input and returns a boolean indicating if the object should be transformed by the data transformer.
The \texttt{Engine} uses the \texttt{Transformer} and \texttt{Filter} to transform the positional arguments, keyword arguments, and results of tasks before being sent to the \texttt{Executor}.
For example, every argument in the positional arguments tuple which passes the filter check is transformed into an identifier using the data transformer.
Each task is encapsulated with a wrapper which will \emph{resolve} any arguments that were replaced with identifiers when the task executes.
The same occurs in reverse for a task's result.

\texttt{Filter} implementations based on object size, pickled object size, and object type are provided.
We initially provide two \texttt{Transformer} implementations: \texttt{PickleFileTransformer} and \texttt{ProxyTransformer}.
The \texttt{PickleFileTransformer} pickles objects and writes the pickled data to a file.
The \texttt{ProxyTransformer} creates proxies of objects using the ProxyStore library~\cite{pauloski2023proxystore,pauloski2024proxystore}.
ProxyStore provides a pass-by-reference like model for distributed Python applications and supports a multitude of communication protocols including DAOS~\cite{hennecke2020daos}, Globus~\cite{foster2011globus,bryce2012saasglobus}, Margo~\cite{py-mochi-margo}, Redis~\cite{redis}, UCX~\cite{UCX-Py}, and ZeroMQ~\cite{hintjens2013zeromq}.

\subsection{Logging and Metrics}

Recording logs and metrics for post-execution analysis is core to any benchmarking framework.
\name{} records the high-level application logs and low-level details of each executed task.
The \texttt{RecordLogger} interface is used to log records of all tasks executed by the \texttt{Engine}.
These records include metrics and metadata of the task, such as the unique task ID, the function name, task IDs of any parent tasks, submission time, completion time, data transformation and resolution times, and execution makespan.
By default, \name{} uses the \texttt{JSONRecordLogger} which logs a JSON representation of the task information to a line-delimited file.
In future work, we would also like to support WfCommon's WfFormat.

\subsection{Task Life-cycle}

An application creates a task by submitting a Python function with corresponding arguments to \texttt{Engine.submit()} which returns a corresponding \texttt{TaskFuture}.
(Applications can also create many tasks by mapping a function onto an iterable of arguments via \texttt{Engine.map()}. For simplicity, we discuss single task submission here, but the same process applies with map.)
The \texttt{Engine} generates a unique ID for the task and wraps the function in a task wrapper.
The \texttt{Transformer} is then applied to the arguments according to the \texttt{Filter}.
Then, the wrapped function and arguments (some or all of which may have been transformed) are passed to the \texttt{Executor} for scheduling and execution.
The \texttt{Executor} returns a future specific to the executor type (e.g., a Globus Compute future for a \texttt{GlobusComputeExecutor}).
This low-level future is then wrapped in a \texttt{TaskFuture}, and the \texttt{TaskFuture} is returned to the client.
If a \texttt{TaskFuture} were passed as input to a task, the \texttt{Engine} will also replace the \texttt{TaskFuture} with the low-level future of the \texttt{Executor}.
This is necessary to ensuring the \texttt{Executor} can schedule the tasks according to the implicit inter-task dependencies.

When a task begins execution, the task wrapper will record information about the execution to propagate back to the \texttt{Engine}.
The task wrapper will also resolve any transformed arguments prior to invoking the original function provided by the client and possibly transform the function result.
The completion of a task (i.e., when the result of the future is set) will trigger a callback which logs all of a task's information and metrics.
If the function result was transformed, the \texttt{TaskFuture} will resolve the result inside of \texttt{TaskFuture.result()}.

\section{Applications}
\label{sec:applications}

We initially provide eight applications within \name{}, summarized in \autoref{tab:applications} and \autoref{fig:app-structures}.
These distributed and parallel applications are diverse, spanning many domains, datasets, and structures to support comprehensive performance evaluation of existing and future systems.

\input{tables/apps}
\input{figures/app-structures}

\subsection{Cholesky Factorization}

Cholesky factorization (also referred to as decomposition) is a fundamental linear algebra operation used in many domains.
The tiled version of Cholesky factorization has been studied extensively, for example, in the context of NUMA machines~\cite{6424759} and from the perspective of communication overhead~\cite{beaumont:hal-03580531}.
The tiled version produces an arbitrarily complex DAG depending on the number of tiles, which makes it a good candidate for evaluating task executors.
The 4$\times$4 tiled DAG is portrayed in~\autoref{fig:app-structures}.

The \texttt{cholesky} application implements a tiled Cholesky factorization which, given an input matrix $A$ that is positive-definite, computes $L$ where $A = L \times L^T$~\cite{jeannot2012cholesky}.
The algorithm comprises four task types: GEMM, a tiled matrix multiplication requiring three inputs; SYRK, a symmetric rank-$k$ update requiring three inputs; TRSM, which solves a triangular matrix equation with two inputs; and POTRF, an untiled Cholesky factorization which operates on a tile of $A$.

The \texttt{cholesky} application takes two user-supplied parameters: $N$, the side length of the input matrix to generate, and $b$, the side length of each square block in the tiled matrix.
As $b$ approaches $N$, the number of blocks in the tiled matrix, and thus the number of tasks required for the factorization, decreases.
Given $B$, a randomly generated $N\times N$ matrix, the positive definite input matrix $A$ is computed by using $A=(B+B^T)+\delta I$, where $\delta=N$ and $I$ is the $N\times N$ identity matrix.

\subsection{Protein Docking}

Protein docking aims to predict the orientation and position of one molecule to another. It is commonly used in structure-based drug design as it helps predict the binding affinity of a ligand (the candidate drug) to the target receptor.
Simulations required to compute docking score are computationally expensive, and the search-space of potential molecules can be expansive.
To improve the time-to-solution, this implementation of protein docking is parallelized and includes ML-in-the-loop.
A model is trained using the results of previous simulations to predict which molecules are most likely to have strong binding scores, thereby significantly reducing the search space.

The \texttt{docking} workflow is based on a reference implementation written in Parsl~\cite{raicu2023parsldock}. The workflow uses Autodock Vina~\cite{trott2010autodock} for the docking simulations and scikit-learn~\cite{scikit-learn}
to construct a KNN-based transformer for the ML model. It is composed of three task types: (1) data preparation, (2) simulation, and (3) ML training and inference.
The workflow has two primary parameters: a CSV file containing the search space of candidate ligands and their associated SMILES strings and a PDBQT file containing the target receptor.
One of the tasks launches a subprocess to execute a \texttt{set-element.tcl} script (provided in the reference implementation) that adds coordinates to the PDB file using VMD~\cite{humphrey1996vmd}, a program used to display and analyse molecular assemblies.

\subsection{Federated Learning}
Federated Learning~(FL) is a paradigm for deep learning across decentralized devices with their own private data.
FL offloads the task of model training to the decentralized devices to avoid communicating their raw training data across the network, providing some level of privacy and reducing data transfer costs.
FL is organized into multiple rounds.
In each round, a central server is responsible for collecting locally-updated model parameters from each device and aggregating the parameters to produce/update a global model.
The new global model is then redistributed to the decentralized devices for further training and the loop repeats for future rounds~\cite{mcmahan2017communication}.

We implement a simple FL application, \texttt{fedlearn}, that simulates a decentralized system with varying number of simulated devices and data distributions.
\texttt{Fedlearn} follows the flow of execution described above and consists of three tasks:
local training, model aggregation, and global model testing.
The first task emulates the local training that is performed on a simulated remote device.
The second task takes the returned locally-trained models for a given round as input to perform a model aggregation step to update the global model.
The third task takes the recently-updated global model and evaluates it using a test dataset that was not used during training.
All tasks are implemented as pure Python functions with model training and evaluation performed using PyTorch~\cite{pytorch}.

The application can be tuned in several ways, including, but not limited to,
the total number of aggregation rounds, 
the number of simulated devices, 
the distribution of data samples across the simulated devices via the Dirichlet distribution,  
training hyperparameters (e.g., epochs, learning rate, minibatch size),
and
fraction of devices randomly sampled to participate in each round.
The application supports four standard deep learning datasets (MNIST~\cite{mnist}, Fashion-MNIST~\cite{fmnist}, CIFAR-10, CIFAR-100~\cite{cifar}), each of which is split into disjoint subsets across each simulated device for local training.
A multi-layer perceptron network with three layers and ReLU activations is used with MNIST and Fashion-MNIST, and a small convolutional neural network with ReLU activations is used with CIFAR-10 and CIFAR-100.

\subsection{MapReduce}

MapReduce~\cite{dean2004mapreduce} is a programming model for parallel big data processing comprised of two tasks types.
Map tasks filter or sort input data, and a reduce task performs a summation operation on the map outputs.
The canonical example for MapReduce is computing words counts in a text corpus.
Here, the map tasks take a subset of documents in the corpus as input and count each word in the subset.
The subset counts are then summed by the reduce task.

The \texttt{mapreduce} application implements this word frequency example.
The goal of this application is to evaluate system responsiveness when processing large datasets.
The implementation can operate in two modes, one in which a text corpus of arbitrary, user-defined size is generated, and another in which user-provided text files can be read.
For a real dataset, we use the publicly available Enron email dataset~\cite{enron}.
Beyond specifying the input corpus or parameters of the randomly generated corpus, the number of mapping tasks and $n$, the number of most frequent words to save, are configurable.

The map task, implemented in Python, takes as input either a string of text or a list of files to read the text string from and returns a \texttt{collections.Counter} object containing the frequencies of each work.
The reduce task takes a list of \texttt{Counter} objects and returns a single \texttt{Counter}.
The application produces an output file containing the $n$ most frequent words and their frequencies.

\subsection{Molecular Design}

Molecules with high ionization potentials (IP) are important for the design of next-generation redox-flow batteries~\cite{ward2021colmena,ward2023colmena}.
Active learning, a process where a surrogate ML model is used to determine which simulations to perform based on previous computations, is commonly employed to efficiently discover high-performing molecules.

The \texttt{moldesign} application is based on a Parsl implementation of ML-guided molecular design~\cite{moldesign}.
The application has three task types.
Simulation tasks compute a molecule's IP, training tasks retrain an ML model based on the results of simulation tasks, and inference tasks use the ML model to predict which molecules will have high IPs and should be simulated.
This application is highly dynamic and does not have strong inter-task dependencies---the client processes task results to determine which new tasks should be submitted.
Molecules are sampled from the open-source QM9 dataset~\cite{ramakrishnan2014quantum}.
The number of initial simulations to perform, simulation batch size, and number of molecules to evaluate in total are configurable.
These parameters control the maximum parallelism of the application and the length of the campaign.

\subsection{Montage}

Montage is a toolkit for creating mosaics from astronomical images~\cite{berriman2003montage}. The Montage Mosaic workflow streamlines the creation of such mosaics by invoking a series of Montage tools on the provided input data. This workflow was adapted from Montage's ``Getting Started" tutorial~\cite{montagetutorial}. 

The \texttt{montage} application is executed using a directory of input images and parameters for the table and header file names.
The 2MASS input images are made available by Montage~\cite{montagedata}.
The application consists of a series of image processing tasks that will (1) reproject the images, (2) update metadata, (3) remove overlaps, and (4) combine images into a mosaic. Parallelism within the workflow occurs during the reprojection of the images, removing overlaps between two images, and removing the background in each input image.
Tasks read and write intermediate files so all workers require access to a shared file system.

\subsection{Failure Injection}

The \texttt{failures} application can inject failures into another \name{} application.
Injecting failures enables analyzing the failure recovery characteristics of executors.
Task-level failure types include runtime exceptions (e.g, divide-by-zero, import error, out-of-memory, open file limit (ulimit) exceeded, and walltime exceeded) and dependency errors from a failed parent task.
System-level failures include task worker, worker manager, and node failures.
The failure type, failure rate, and base application to inject failures into are configurable.

\subsection{Synthetic Workflow}

The \texttt{synthetic} application is used to create synthetic computational workflows and is useful for stress testing systems.
Tasks in this application are no-op sleep tasks which take in some random data and, optionally, produce some random data.
One of four structures for the workflow DAG can be chosen: sequential, reduce, bag-of-tasks, and diamond, as described in \autoref{fig:app-structures}.
The number of tasks, input and output data sizes, and sleep times are all configurable.

\section{Evaluation}
\label{sec:evaluation}

We showcase the kinds of performance evaluations possible with \name{} using the provided applications.
We draw some general conclusions but do not make an exhaustive comparison between executors.
Rather, we aim to demonstrate the varied performance characteristics of our supported applications and plugins, highlight the kinds of investigations or analyses that can be performed with \name{}, and pose interesting questions for future investigations.
We use a \texttt{compute-zen-3} node, with two 64-core CPUs and 256 GB memory, on Chameleon Cloud’s CHI@TACC cluster for evaluation~\cite{keahey2020lessons}.

\subsection{Application Makespan}
\label{sec:evaluation:makespan}

\input{figures/single-node-makespans}
\input{tables/app-executions}

We first compare application makespan, which includes executor and worker initialization, application execution, and shutdown, across each task executor.
The space of possible configurations for each application and executor is combinatorially explosive.
Thus, we choose application parameters, where possible, which result in high numbers of short tasks to accentuate the effects of overheads in the respective executors.
Parameters are summarized in \autoref{tab:app-execution}.
We also prefer configurations which reduce run-to-run variances, except for \texttt{docking} which is inherently stochastic.
For each executor, we use the respective equivalent of a default local/single-node deployment, but we note that it is reasonable to expect performance improvements by tuning each executor deployment to the specific application and hardware.

The results, presented in \autoref{fig:makespans}, indicate that no executor is optimal and lead us to ask further questions.
Why are the following 2--3$\times$ faster than the others: Ray in \texttt{cholesky}, Dask and Parsl in \texttt{moldesign}, and Dask in \texttt{montage}?
How does performance correlate to average task duration or data flow volume?
How do different executors deal with nested parallelism (i.e., tasks which invoke multi-threaded code)?

We observe that Dask performs the best in applications with small maximum object sizes, such as \texttt{docking}, \texttt{moldesign}, and \texttt{montage} where, as shown in \autoref{tab:app-execution}, the maximum serialized object sizes are less than $\sim$1~MB.
However, Dask is slow with applications that embed large objects in the task graph, such as the 114~MB mapper outputs in \texttt{mapreduce}.
Ray marks input arrays as immutable enabling optimizations which yield considerable speedups in \texttt{cholesky}.
Applications with nested parallelism (the simulation codes in \texttt{docking} and \texttt{moldesign} and tensor operations in \texttt{fedlearn}) lead to different outcomes.
Globus Compute, Parsl, and \texttt{ProcessPoolExecutor} required setting \texttt{OMP\_NUM\_THREADS=1} to prevent resource contention leading to applications hanging, whereas Dask, Ray, and TaskVine worked immediately with all task types, albeit with varied performance.
The Globus Compute service limits task payloads to 10~MB so the \texttt{cholesky}, \texttt{fedlearn}, and \texttt{mapreduce} applications are not natively supported and necessitate alternative data management systems (discussed further in \autoref{sec:evaluation:data}).

\subsection{Scaling Performance}
\label{sec:evaluation:scaling}

We evaluate scaling performance of each executor using the \texttt{synthetic} app by executing \num{1000} no-op, no-data tasks and recording the task completion rate as a function of the number of workers on a single node.
Here, the client submits $n$ initial tasks where $n$ is the number of workers and submits new tasks as running tasks complete.
This configuration is intended to stress-test all aspects of the system including scheduler throughput, worker overheads, and client task result latency.
We disable task result caching where applicable.

\input{figures/scaling-performance}

The results are presented in \autoref{fig:scaling}.
The \texttt{Process\-PoolExecutor} performs the best because, unlike the other executors, there is no scheduler.
Thus, this serves as a good baseline for this single-node scaling setup;
however, the lack of scheduler also means the \texttt{ProcessPoolExecutor} lacks features useful for optimizing real applications such as multi-node support, data-aware task placement, and result caching.
The general trend for Dask, Ray, and TaskVine is similar; task throughput increases up to four or eight workers and then degrades at high worker counts.
However, Ray and TaskVine are both faster, with Ray being 5--10$\times$ faster than Dask.
This can, in part, be attributed to Dask being pure Python while TaskVine's core is C and Ray's core is C++.
Parsl, which is pure Python, exhibits superior scaling efficiency, closing the performance gap to Ray at larger scales.
Globus Compute's task throughput is limited by its cloud service, but we do observe strong scaling performance with more workers as task requests and results can be more efficiently batched which amortizes cloud overheads.

\subsection{Data Transfer}
\label{sec:evaluation:data}

\input{figures/data-transfer}

We examine the effects of data transfer on task latency and evaluate the \texttt{Transformer} plugins in \autoref{fig:data-transfer}.
We submit tasks to a pool of 32 workers and measure the average round-trip task time using the \texttt{synthetic} application.
The client generates $b$ bytes of random data as input to the task and the task returns $b$ bytes of random data.
We compare the baseline performance of the executors to using two different transformers: \texttt{PickleFileTransformer}, which writes pickled task data to the local NVMe drive, and ProxyStore, which we configured to use a Redis server to store intermediate data.

Dask and Parsl exhibit similar behaviour with task payloads greater than 100~kB inducing considerable increases in task latency.
Using an alternate mechanism for data transfer alleviates much of this overhead, leading to 5.8$\times$ and 4.4$\times$ speedups for Dask and Parsl, respectively, at the largest data sizes.
Globus Compute benefits the most from alternative data transfer mechanisms such as ProxyStore because the baseline method relies on data transfer to/from the cloud which is considerably slower. Use of ProxyStore also avoids Globus Compute's 10~MB task payload limit.
The \texttt{ProcessPoolExecutor}, due to its simplicity, does not benefit much from either alternative transfer mechanisms.
Ray and TaskVine perform well in all scenarios because Ray uses a distributed object store for large task data and TaskVine communicates intermediate data by files. Thus, these systems already employ techniques similar to the data transformers we evaluated.

\section{Conclusion}
\label{sec:conclusion}

We have proposed \name{}, a performance evaluation platform for task-based execution frameworks.
\name{} aims to provide a standard system for benchmarking frameworks.
Benchmarking applications can be written in a framework agnostic manner then evaluated using \name{}' extensive plugin system.
\name{} provides many reference applications, a diverse set of supported task executors and data management systems, and performance and metadata logging.
We then showcased \name{} through a survey of evaluations to understand performance characteristics of the applications and executors, such as task overheads, data management, and scalability.
Our hope is that \name{} will be a long-standing tool used to provide a common ground for evaluation and to facilitate the advancement in the state-of-the-art for parallel application execution.

\section*{Acknowledgments}
We acknowledge the helpful support and guidance from the Cooperative Computing Lab at the University of Notre Dame with integrating TaskVine.
This research was supported in part by Argonne National Laboratory under U.S. Department of Energy under Contract DE-AC02-06CH1135 and by the National Science Foundation under Grant 2004894 and 2209919
Results were obtained using the Chameleon testbed supported by the National Science Foundation.

{
\scriptsize
\balance
\bibliographystyle{IEEEtran}
\bibliography{refs}
}

\end{document}

%% file: figures/framework-stack.tex
\begin{figure}
    \centering
    \includegraphics[width=\columnwidth,trim={6mm 0px 6mm 0px},clip]{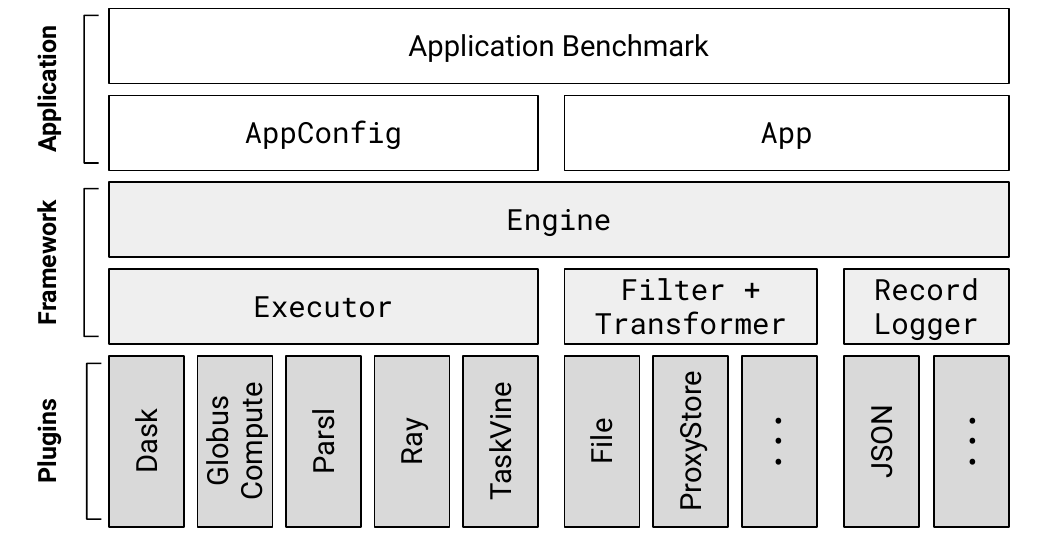}
    \caption{Overview of the \name{} stack.}
    \label{fig:framework-stack}
\end{figure}

%% file: listings/application.tex
\begin{lstlisting}[
    style=PythonStyle,
    label={lst:application},
    caption={Example application structure within \name{}.},
    float,
    floatplacement=bt
]
@register('app')
class FooAppConfig(AppConfig):
    name: str = 'foo'
    sleep: float = Field(description='...')
    count: int = Field(1, description='...')

    def get_app(self) -> FooApp: ...

class FooApp:
    def __init__(self, ...) -> None: ...

    def run(self, engine: Engine, run_dir: Path) -> None:
        ...

    def close(self) -> None: ...
\end{lstlisting}

%% file: tables/engines.tex
\begin{table*}
\centering
\caption{Overview of the execution engines supported within \name{}.}
\label{tab:engines}
\ifpreprint
    \footnotesize
\else
    \scriptsize
\fi
\begin{tabular}{lclccccc}
\toprule
                             &                               &                   & \multicolumn{3}{c}{Scheduler}    & \multicolumn{2}{c}{Deployment} \\
\cmidrule(lr){4-6} \cmidrule{7-8}
Name                         & Reference                     & Languages         & Distributed  & Dataflow   & Locality-Aware & Distributed  & Batch Systems \\
\midrule
\texttt{ThreadPoolExecutor}  & \cite{concurrentfutres}       & Python            &              & & & & \\
\texttt{ProcessPoolExecutor} & \cite{concurrentfutres}       & Python            &              & & & & \\
Dask Distributed             & \cite{rocklin2015dask}        & Python            &              & $\checkmark$ & $\checkmark$ & $\checkmark$ & $\checkmark$ \\
Globus Compute               & \cite{chard20funcx}           & Python            &              &              &              & $\checkmark$ & $\checkmark$ \\
Parsl                        & \cite{babuji19parsl}          & Python            &              & $\checkmark$ &              & $\checkmark$ & $\checkmark$ \\
Ray                          & \cite{moritz2018ray}          & C++, Java, Python & $\checkmark$ & $\checkmark$ & $\checkmark$ & $\checkmark$ & $\checkmark$ \\
TaskVine                     & \cite{slydelgado2023taskvine} & C, Python       &              & $\checkmark$ & $\checkmark$ & $\checkmark$ & $\checkmark$ \\
\bottomrule
\end{tabular}
\end{table*}

%% file: tables/apps.tex
\begin{table*}
\centering
\caption{Overview of the applications implemented within \name{}.}
\label{tab:applications}
\scriptsize
\begin{tabular}{lcllll}
\toprule
Name & Reference & Domain & Dataset(s) & Task Types(s) & Data Format(s) \\
\midrule
\texttt{cholesky}  & \cite{jeannot2012cholesky}    & Linear Algebra   & Randomly Generated            & Python Functions & In-memory \\
\texttt{docking}   & \cite{raicu2023parsldock}     & Drug Discovery   & C-ABL Kinase Domain~\cite{autodock-examples}, Zinc Ord. Compounds~\cite{clyde2023ai}                      & Executable, Python Functions & File \\
\texttt{fedlearn}  & \cite{flox} & Machine Learning & MNIST~\cite{mnist}, FEMNIST~\cite{fmnist}, CIFAR-10/100~\cite{cifar}                      & Python Functions & In-memory \\

\texttt{mapreduce} & \cite{dean2004mapreduce}      & Text Analysis    & Randomly Generated, Enron Corpus~\cite{enron} & Python Functions & File, In-memory \\
\texttt{moldesign} & \cite{moldesign}              & Molecular Design & QM9~\cite{ramakrishnan2014quantum}            & Python Functions & In-memory \\
\texttt{montage}   & \cite{montage}                & Astronomy        & Montage Images~\cite{montage} & Executable & File \\
\texttt{failures}  & ---                           & ---              & ---  & Executable, Python Functions & File, In-memory \\
\texttt{synthetic} & \cite{pauloski2023proxystore} & ---              & Randomly Generated            & Python Functions & In-memory \\
\bottomrule
\end{tabular}
\end{table*}

%% file: figures/app-structures.tex
\begin{figure*}
    \centering
    \includegraphics[width=0.9\textwidth,trim={0 4px 0 4px},clip]{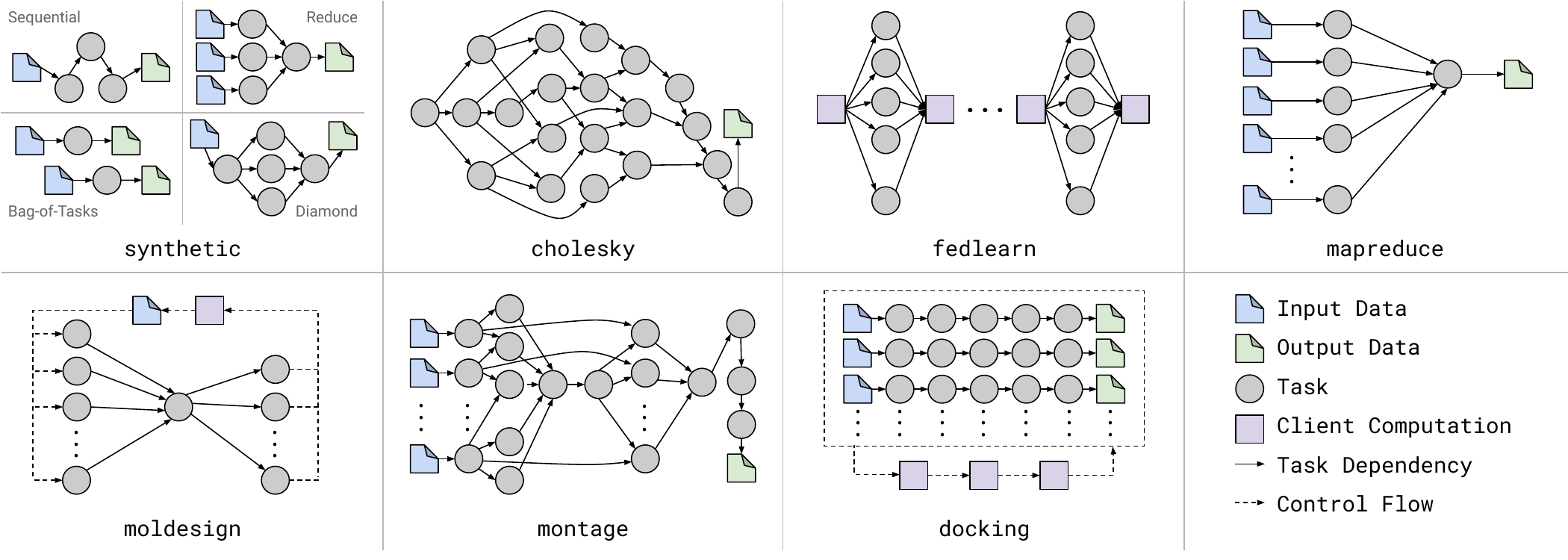}
    \caption{Example task dependency diagrams for each application. In most applications, the exact structure depends on the application configuration.}
    \label{fig:app-structures}
\end{figure*}

%% file: figures/single-node-makespans.tex
\begin{figure*}
    \centering
    \includegraphics[width=\textwidth,trim={0 2pt 0 2pt},clip]{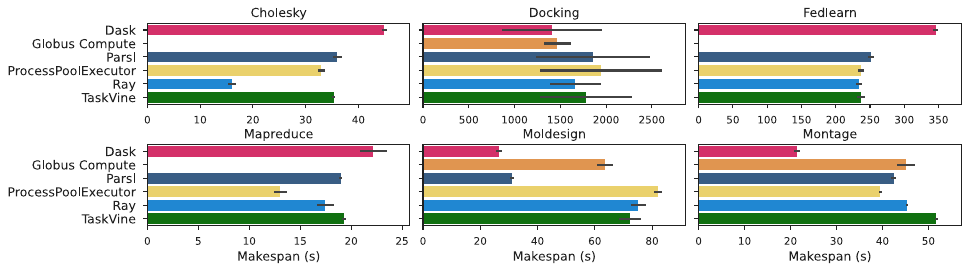}
    \caption{Average application makespan over three runs. Error bars denote standard deviation.}
    \label{fig:makespans}
\end{figure*}

%% file: tables/app-executions.tex
\begin{table*}
\centering
\caption{Summary of application configurations used in \autoref{fig:makespans}.}
\label{tab:app-execution}
\ifpreprint
    \footnotesize
\else
    \scriptsize
\fi
\begin{tabular}{lcccl}
\toprule
Application        & Workers & Task Count & Max Serialized Object Size & Parameters \\
\midrule
\texttt{cholesky}  & 64      & 385 & 24~MB     & Matrix Size: \num{10000}$\times$\num{10000}, Block Size: \num{1000}$\times$\num{1000}\\
\texttt{docking}   & 32      & 192 & $O(1)$~kB & Initial Simulations: 3, Batch Size: 8, Rounds: 3 \\
\texttt{fedlearn}  & 32      & 48  & 20~MB     & Dataset: MNIST, Clients: 16, Batch Size: 32, Rounds: 3, Epochs/Round: 1\\
\texttt{mapreduce} & 32      & 33  & 114~MB    & Dataset: Enron Email Corpus, Map Task Count: 32\\
\texttt{moldesign} & 32      & 346 & $O(1)$~MB & Initial Simulations: 16, Batch Size: 16, Search Count: 64 \\
\texttt{montage}   & 32      & 419 & $O(1)$~kB & --- \\
\bottomrule
\end{tabular}
\end{table*}

%% file: figures/scaling-performance.tex
\begin{figure}
    \centering
    \includegraphics[width=\columnwidth,trim={0 2pt 0 2pt},clip]{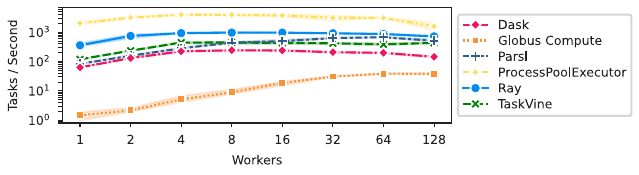}
    \caption{Executor scaling performance with no-op tasks. Each configuration is repeated three times and shaded regions represent the standard deviation.}
    \label{fig:scaling}
\end{figure}

%% file: figures/data-transfer.tex
\begin{figure}
    \centering
    \includegraphics[width=\columnwidth,trim={0 2pt 0 2pt},clip]{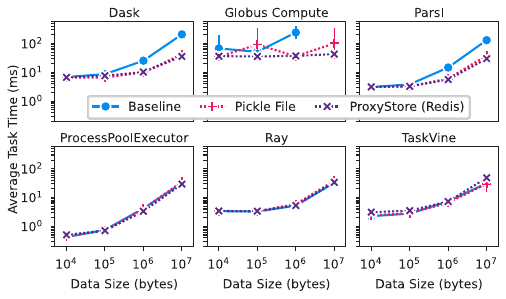}
    \caption{Average round-trip time for no-op tasks as a function of input/output data size. Error bars denote standard deviation from three runs of 320 tasks ($10\times 32$ workers).
    The Globus Compute baseline is not evaluated at 10~MB due to task payload limits of the Globus Compute service.
    }
    \label{fig:data-transfer}
\end{figure}

%% file: main.bbl
\begin{thebibliography}{10}
\providecommand{\url}[1]{#1}
\csname url@samestyle\endcsname
\providecommand{\newblock}{\relax}
\providecommand{\bibinfo}[2]{#2}
\providecommand{\BIBentrySTDinterwordspacing}{\spaceskip=0pt\relax}
\providecommand{\BIBentryALTinterwordstretchfactor}{4}
\providecommand{\BIBentryALTinterwordspacing}{\spaceskip=\fontdimen2\font plus
\BIBentryALTinterwordstretchfactor\fontdimen3\font minus \fontdimen4\font\relax}
\providecommand{\BIBforeignlanguage}[2]{{%
\expandafter\ifx\csname l@#1\endcsname\relax
\typeout{** WARNING: IEEEtran.bst: No hyphenation pattern has been}%
\typeout{** loaded for the language `#1'. Using the pattern for}%
\typeout{** the default language instead.}%
\else
\language=\csname l@#1\endcsname
\fi
#2}}
\providecommand{\BIBdecl}{\relax}
\BIBdecl

\bibitem{rocklin2015dask}
M.~Rocklin, ``Dask: Parallel computation with blocked algorithms and task scheduling,'' in \emph{14th Python in Science Conference}, vol. 130, 2015, p. 136.

\bibitem{babuji19parsl}
\BIBentryALTinterwordspacing
Y.~Babuji, A.~Woodard, Z.~Li, D.~S. Katz, B.~Clifford, R.~Kumar, L.~Lacinski, R.~Chard, J.~M. Wozniak, I.~Foster, M.~Wilde, and K.~Chard, ``Parsl: Pervasive parallel programming in {P}ython,'' in \emph{28th ACM International Symposium on High-Performance Parallel and Distributed Computing}, 2019. [Online]. Available: \url{https://doi.org/10.1145/3307681.3325400}
\BIBentrySTDinterwordspacing

\bibitem{moritz2018ray}
P.~Moritz, R.~Nishihara, S.~Wang, A.~Tumanov, R.~Liaw, E.~Liang, M.~Elibol, Z.~Yang, W.~Paul, M.~I. Jordan, and I.~Stoica, ``{Ray}: A distributed framework for emerging {AI} applications,'' in \emph{Proceedings of the 13th USENIX Conference on Operating Systems Design and Implementation}, ser. OSDI'18.\hskip 1em plus 0.5em minus 0.4em\relax USA: USENIX Association, 2018, p. 561–577.

\bibitem{dongarra1988linpack}
J.~J. Dongarra, ``The {LINPACK} benchmark: An explanation,'' in \emph{1st International Conference on Supercomputing}.\hskip 1em plus 0.5em minus 0.4em\relax Berlin, Heidelberg: Springer-Verlag, 1988, p. 456–474.

\bibitem{tpc}
``{Transaction Processing Performance Council},'' \url{https://www.tpc.org/}. Accessed May 2024.

\bibitem{unixbench}
``{UnixBench},'' \url{https://github.com/kdlucas/byte-unixbench}. Accessed May 2024.

\bibitem{mattson2020mlperf}
P.~Mattson, C.~Cheng, G.~Diamos, C.~Coleman, P.~Micikevicius, D.~Patterson, H.~Tang, G.-Y. Wei, P.~Bailis, V.~Bittorf, D.~Brooks, D.~Chen, D.~Dutta, U.~Gupta, K.~Hazelwood, A.~Hock, X.~Huang, A.~Ike, B.~Jia, D.~Kang, D.~Kanter, N.~Kumar, J.~Liao, G.~Ma, D.~Narayanan, T.~Oguntebi, G.~Pekhimenko, L.~Pentecost, V.~J. Reddi, T.~Robie, T.~S. John, T.~Tabaru, C.-J. Wu, L.~Xu, M.~Yamazaki, C.~Young, and M.~Zaharia, ``{MLPerf training benchmark},'' \emph{Proceedings of Machine Learning and Systems}, vol.~2, pp. 336--349, 2020.

\bibitem{reddi2020mlperf}
V.~J. Reddi, C.~Cheng, D.~Kanter, P.~Mattson, G.~Schmuelling, C.-J. Wu, B.~Anderson, M.~Breughe, M.~Charlebois, W.~Chou, mesh Chukka, C.~Coleman, S.~Davis, P.~Deng, G.~Diamos, J.~Duke, D.~Fick, J.~S. Gardner, I.~Hubara, S.~Idgunji, T.~B. Jablin, J.~Jiao, T.~S. John, P.~Kanwar, D.~Lee, J.~Liao, A.~Lokhmotov, F.~Massa, P.~Meng, P.~Micikevicius, C.~Osborne, G.~Pekhimenko, A.~T.~R. Rajan, D.~Sequeira, A.~Sirasao, F.~Sun, H.~Tang, M.~Thomson, F.~Wei, E.~Wu, L.~Xu, K.~Yamada, B.~Yu, G.~Yuan, A.~Zhong, P.~Zhang, and Y.~Zhou, ``{MLPerf Inference Benchmark},'' in \emph{ACM/IEEE 47th Annual International Symposium on Computer Architecture}.\hskip 1em plus 0.5em minus 0.4em\relax IEEE, 2020, pp. 446--459.

\bibitem{paperswithcode}
``{Papers with Code},'' \url{https://paperswithcode.com/datasets}. Accessed May 2024.

\bibitem{bailey1995parallel}
D.~Bailey, T.~Harris, W.~Saphir, R.~Van Der~Wijngaart, A.~Woo, and M.~Yarrow, ``The {NAS} parallel benchmarks 2.0,'' Technical Report NAS-95-020, NASA Ames Research Center, Tech. Rep., 1995.

\bibitem{montage}
``Montage: An astronomical image mosaic engine,'' \url{http://montage.ipac.caltech.edu/}. Accessed Mar. 2024.

\bibitem{deelman2008montageworkflow}
E.~Deelman, G.~Singh, M.~Livny, B.~Berriman, and J.~Good, ``{The cost of doing science on the cloud: The Montage example},'' in \emph{SC'08: ACM/IEEE Conference on Supercomputing}, 2008, pp. 1--12.

\bibitem{colman2022wfcommons}
\BIBentryALTinterwordspacing
T.~Coleman, H.~Casanova, L.~Pottier, M.~Kaushik, E.~Deelman, and R.~{Ferreira da Silva}, ``{WfCommons}: A framework for enabling scientific workflow research and development,'' \emph{Future Generation Computer Systems}, vol. 128, pp. 16--27, 2022. [Online]. Available: \url{https://www.sciencedirect.com/science/article/pii/S0167739X21003897}
\BIBentrySTDinterwordspacing

\bibitem{chard20funcx}
\BIBentryALTinterwordspacing
R.~Chard, Y.~Babuji, Z.~Li, T.~Skluzacek, A.~Woodard, B.~Blaiszik, I.~Foster, and K.~Chard, ``{func{X}: A Federated Function Serving Fabric for Science},'' in \emph{29th International Symposium on High-Performance Parallel and Distributed Computing}.\hskip 1em plus 0.5em minus 0.4em\relax ACM, 2020. [Online]. Available: \url{http://dx.doi.org/10.1145/3369583.3392683}
\BIBentrySTDinterwordspacing

\bibitem{alsaadi2021radical}
A.~Alsaadi, L.~Ward, A.~Merzky, K.~Chard, I.~Foster, S.~Jha, and M.~Turilli, ``Radical-{P}ilot and {P}arsl: Executing heterogeneous workflows on {HPC} platforms,'' \emph{arXiv preprint arXiv:2105.13185}, 2021.

\bibitem{airflow}
``{Apache Airflow},'' \url{https://airflow.apache.org/}. Accessed May 2024.

\bibitem{jain2015fireworks}
A.~Jain, S.~P. Ong, W.~Chen, B.~Medasani, X.~Qu, M.~Kocher, M.~Brafman, G.~Petretto, G.-M. Rignanese, G.~Hautier, D.~Gunter, and K.~A. Persson, ``{FireWorks}: A dynamic workflow system designed for high-throughput applications,'' \emph{Concurrency and Computation: Practice and Experience}, vol.~27, no.~17, pp. 5037--5059, 2015.

\bibitem{albrecht2012makeflow}
\BIBentryALTinterwordspacing
M.~Albrecht, P.~Donnelly, P.~Bui, and D.~Thain, ``Makeflow: A portable abstraction for data intensive computing on clusters, clouds, and grids,'' in \emph{Proceedings of the 1st ACM SIGMOD Workshop on Scalable Workflow Execution Engines and Technologies}, ser. SWEET '12.\hskip 1em plus 0.5em minus 0.4em\relax New York, NY, USA: Association for Computing Machinery, 2012. [Online]. Available: \url{https://doi.org/10.1145/2443416.2443417}
\BIBentrySTDinterwordspacing

\bibitem{di2017nextflow}
P.~Di~Tommaso, M.~Chatzou, E.~W. Floden, P.~P. Barja, E.~Palumbo, and C.~Notredame, ``Nextflow enables reproducible computational workflows,'' \emph{Nature biotechnology}, vol.~35, no.~4, pp. 316--319, 2017.

\bibitem{deelman15pegasus}
\BIBentryALTinterwordspacing
E.~Deelman, K.~Vahi, G.~Juve, M.~Rynge, S.~Callaghan, P.~J. Maechling, R.~Mayani, W.~Chen, R.~{Ferreira da Silva}, M.~Livny, and K.~Wenger, ``Pegasus, a workflow management system for science automation,'' \emph{Future Generation Computer Systems}, vol.~46, pp. 17--35, 2015. [Online]. Available: \url{https://www.sciencedirect.com/science/article/pii/S0167739X14002015}
\BIBentrySTDinterwordspacing

\bibitem{wilde11swift}
\BIBentryALTinterwordspacing
M.~Wilde, M.~Hategan, J.~M. Wozniak, B.~Clifford, D.~S. Katz, and I.~Foster, ``Swift: A language for distributed parallel scripting,'' \emph{Parallel Computing}, vol.~37, no.~9, pp. 633--652, 2011. [Online]. Available: \url{https://www.sciencedirect.com/science/article/pii/S0167819111000524}
\BIBentrySTDinterwordspacing

\bibitem{wozniak13swiftt}
J.~M. {Wozniak}, T.~G. {Armstrong}, M.~{Wilde}, D.~S. {Katz}, E.~{Lusk}, and I.~T. {Foster}, ``{Swift/T}: Large-scale application composition via distributed-memory dataflow processing,'' in \emph{13th IEEE/ACM International Symposium on Cluster, Cloud, and Grid Computing}, 2013, pp. 95--102.

\bibitem{slydelgado2023taskvine}
\BIBentryALTinterwordspacing
B.~Sly-Delgado, T.~S. Phung, C.~Thomas, D.~Simonetti, A.~Hennessee, B.~Tovar, and D.~Thain, ``{TaskVine}: Managing in-cluster storage for high-throughput data intensive workflows,'' in \emph{Proceedings of the SC '23 Workshops of The International Conference on High Performance Computing, Network, Storage, and Analysis}, ser. SC-W '23.\hskip 1em plus 0.5em minus 0.4em\relax New York, NY, USA: Association for Computing Machinery, 2023, p. 1978–1988. [Online]. Available: \url{https://doi.org/10.1145/3624062.3624277}
\BIBentrySTDinterwordspacing

\bibitem{pegasus-examples}
``Pegasus examples,'' \url{https://github.com/pegasus-isi/ACCESS-Pegasus-Examples/}. Accessed May 2024.

\bibitem{dask-benchmarks}
``Dask benchmarks,'' \url{https://github.com/dask/dask-benchmarks}. Accessed May 2024.

\bibitem{workflowhub}
\BIBentryALTinterwordspacing
C.~Goble, S.~Soiland-Reyes, F.~Bacall, S.~Owen, A.~Williams, I.~Eguinoa, B.~Droesbeke, S.~Leo, L.~Pireddu, L.~Rodríguez-Navas, J.~M. Fernández, S.~Capella-Gutierrez, H.~Ménager, B.~Grüning, B.~Serrano-Solano, P.~Ewels, and F.~Coppens, ``Implementing {FAIR} digital objects in the {EOSC}-life workflow collaboratory,'' 2021. [Online]. Available: \url{https://doi.org/10.5281/zenodo.4605654}
\BIBentrySTDinterwordspacing

\bibitem{silva2014community}
R.~Ferreira~da Silva, W.~Chen, G.~Juve, K.~Vahi, and E.~Deelman, ``Community resources for enabling research in distributed scientific workflows,'' in \emph{IEEE 10th International Conference on eScience, eScience 2014}, vol.~1, 10 2014.

\bibitem{casanova2020wrench}
H.~Casanova, R.~Ferreira~da Silva, R.~Tanaka, S.~Pandey, G.~Jethwani, W.~Koch, S.~Albrecht, J.~Oeth, and F.~Suter, ``Developing accurate and scalable simulators of production workflow management systems with {WRENCH},'' \emph{Future Generation Computer Systems}, vol. 112, pp. 162--175, 2020.

\bibitem{casanova2014simgrid}
\BIBentryALTinterwordspacing
H.~Casanova, A.~Giersch, A.~Legrand, M.~Quinson, and F.~Suter, ``{Versatile, Scalable, and Accurate Simulation of Distributed Applications and Platforms},'' \emph{Journal of Parallel and Distributed Computing}, vol.~74, no.~10, pp. 2899--2917, Jun. 2014. [Online]. Available: \url{http://hal.inria.fr/hal-01017319}
\BIBentrySTDinterwordspacing

\bibitem{katz2016skeleton}
\BIBentryALTinterwordspacing
D.~S. Katz, A.~Merzky, Z.~Zhang, and S.~Jha, ``Application skeletons: Construction and use in escience,'' \emph{Future Generation Computer Systems}, vol.~59, pp. 114--124, 2016. [Online]. Available: \url{https://www.sciencedirect.com/science/article/pii/S0167739X15003143}
\BIBentrySTDinterwordspacing

\bibitem{jeongchul2019functionbench}
J.~Kim and K.~Lee, ``{FunctionBench}: A suite of workloads for serverless cloud function service,'' in \emph{IEEE 12th International Conference on Cloud Computing (CLOUD)}, 2019, pp. 502--504.

\bibitem{pascal2020faasdom}
\BIBentryALTinterwordspacing
P.~Maissen, P.~Felber, P.~Kropf, and V.~Schiavoni, ``{FaaSdom}: A benchmark suite for serverless computing,'' in \emph{14th ACM International Conference on Distributed and Event-Based Systems}, ser. DEBS '20.\hskip 1em plus 0.5em minus 0.4em\relax New York, NY, USA: Association for Computing Machinery, 2020, p. 73–84. [Online]. Available: \url{https://doi.org/10.1145/3401025.3401738}
\BIBentrySTDinterwordspacing

\bibitem{copik2021sebs}
M.~Copik, G.~Kwasniewski, M.~Besta, M.~Podstawski, and T.~Hoefler, ``{SeBS}: A serverless benchmark suite for function-as-a-service computing,'' in \emph{Proceedings of the 22nd International Middleware Conference}, 2021, pp. 64--78.

\bibitem{concurrentfutres}
``{Python Concurrent Execution},'' \url{https://docs.python.org/3/library/concurrent.futures.html}. Accessed May 2024.

\bibitem{pauloski2023proxystore}
\BIBentryALTinterwordspacing
J.~G. Pauloski, V.~Hayot-Sasson, L.~Ward, N.~Hudson, C.~Sabino, M.~Baughman, K.~Chard, and I.~Foster, ``{Accelerating Communications in Federated Applications with Transparent Object Proxies},'' in \emph{Proceedings of the International Conference for High Performance Computing, Networking, Storage and Analysis}, ser. SC '23, New York, NY, USA, 2023. [Online]. Available: \url{https://doi.org/10.1145/3581784.3607047}
\BIBentrySTDinterwordspacing

\bibitem{pauloski2024proxystore}
\BIBentryALTinterwordspacing
J.~G. Pauloski, V.~Hayot-Sasson, L.~Ward, A.~Brace, A.~Bauer, K.~Chard, and I.~Foster, ``{O}bject {P}roxy {P}atterns for {A}ccelerating {D}istributed {A}pplications,'' 2024. [Online]. Available: \url{https://arxiv.org/abs/2407.01764}
\BIBentrySTDinterwordspacing

\bibitem{hennecke2020daos}
M.~Hennecke, ``{DAOS}: A scale-out high performance storage stack for storage class memory,'' \emph{Supercomputing frontiers}, vol.~40, 2020.

\bibitem{foster2011globus}
I.~Foster, ``{Globus Online}: Accelerating and democratizing science through cloud-based services,'' \emph{IEEE Internet Computing}, vol.~15, no.~3, pp. 70--73, 2011.

\bibitem{bryce2012saasglobus}
\BIBentryALTinterwordspacing
B.~Allen, J.~Bresnahan, L.~Childers, I.~Foster, G.~Kandaswamy, R.~Kettimuthu, J.~Kordas, M.~Link, S.~Martin, K.~Pickett, and S.~Tuecke, ``Software as a service for data scientists,'' \emph{Communications of the ACM}, vol.~55, no.~2, p. 81–88, feb 2012. [Online]. Available: \url{https://doi.org/10.1145/2076450.2076468}
\BIBentrySTDinterwordspacing

\bibitem{py-mochi-margo}
``{Py-Margo},'' \url{https://github.com/mochi-hpc/py-mochi-margo}. Accessed Mar 2023.

\bibitem{redis}
``Redis,'' 2023, \url{https://redis.io/}. Accessed Mar 2023.

\bibitem{UCX-Py}
``{UCX-Py},'' \url{https://ucx-py.readthedocs.io/en/latest/}. Accessed Mar 2023.

\bibitem{hintjens2013zeromq}
P.~Hintjens, \emph{Zero{MQ}: Messaging for Many Applications}.\hskip 1em plus 0.5em minus 0.4em\relax O'Reilly Media, Inc., 2013.

\bibitem{jeannot2012cholesky}
E.~Jeannot, ``Performance analysis and optimization of the tiled {C}holesky factorization on {NUMA} machines,'' in \emph{5th International Symposium on Parallel Architectures, Algorithms and Programming}, 2012, pp. 210--217.

\bibitem{raicu2023parsldock}
J.~Raicu, V.~Hayot-Sasson, K.~Chard, and I.~Foster, ``Navigating the molecular maze: A {P}ython-powered approach to virtual drug screening,'' 2023, \url{https://github.com/Parsl/parsl-docking-tutorial}. Accessed May 2024.

\bibitem{autodock-examples}
``{AutoDock Vina: P}ython scripting,'' \url{https://github.com/ccsb-scripps/AutoDock-Vina/tree/develop/example/python_scripting}. Accessed May 2024.

\bibitem{clyde2023ai}
A.~Clyde, X.~Liu, T.~Brettin, H.~Yoo, A.~Partin, Y.~Babuji, B.~Blaiszik, J.~Mohd-Yusof, A.~Merzky, M.~Turilli \emph{et~al.}, ``{AI}-accelerated protein-ligand docking for {SARS-CoV-2} is 100-fold faster with no significant change in detection,'' \emph{Scientific {R}eports}, vol.~13, no.~1, p. 2105, 2023.

\bibitem{flox}
N.~Kotsehub, M.~Baughman, R.~Chard, N.~Hudson, P.~Patros, O.~Rana, I.~Foster, and K.~Chard, ``{FLoX}: Federated learning with {FaaS} at the edge,'' in \emph{IEEE 18th International Conference on e-Science}.\hskip 1em plus 0.5em minus 0.4em\relax IEEE, 2022, pp. 11--20.

\bibitem{mnist}
L.~Deng, ``The {MNIST} database of handwritten digit images for machine learning research,'' \emph{IEEE Signal Processing Magazine}, vol.~29, no.~6, pp. 141--142, 2012.

\bibitem{fmnist}
H.~Xiao, K.~Rasul, and R.~Vollgraf, ``{Fashion-MNIST}: A novel image dataset for benchmarking machine learning algorithms,'' \emph{arXiv:1708.07747}, 2017.

\bibitem{cifar}
A.~Krizhevsky, G.~Hinton \emph{et~al.}, ``Learning multiple layers of features from tiny images,'' 2009.

\bibitem{dean2004mapreduce}
\BIBentryALTinterwordspacing
J.~Dean and S.~Ghemawat, ``{MapReduce}: Simplified data processing on large clusters,'' in \emph{6th Symposium on Operating Systems Design \& Implementation (OSDI 04)}.\hskip 1em plus 0.5em minus 0.4em\relax San Francisco, CA: USENIX Association, Dec. 2004. [Online]. Available: \url{https://www.usenix.org/conference/osdi-04/mapreduce-simplified-data-processing-large-clusters}
\BIBentrySTDinterwordspacing

\bibitem{enron}
``{Enron Email Corpus},'' \url{https://www.cs.cmu.edu/~enron/}. Accessed May 2024.

\bibitem{moldesign}
``Molecular design in {P}arsl,'' \url{https://github.com/ExaWorks/molecular-design-parsl-demo}. Accessed May 2024.

\bibitem{ramakrishnan2014quantum}
R.~Ramakrishnan, P.~O. Dral, M.~Rupp, and O.~A. von Lilienfeld, ``Quantum chemistry structures and properties of 134 kilo molecules,'' \emph{Scientific Data}, vol.~1, 2014.

\bibitem{6424759}
E.~Jeannot, ``Performance analysis and optimization of the tiled {C}holesky factorization on {NUMA} machines,'' in \emph{5th International Symposium on Parallel Architectures, Algorithms and Programming}, 2012, pp. 210--217.

\bibitem{beaumont:hal-03580531}
\BIBentryALTinterwordspacing
O.~Beaumont, L.~Eyraud-Dubois, M.~V{\'e}rit{\'e}, and J.~Langou, ``{I/O-Optimal Algorithms for Symmetric Linear Algebra Kernels},'' in \emph{{ACM Symposium on Parallelism in Algorithms and Architectures}}, 2022. [Online]. Available: \url{https://hal.inria.fr/hal-03580531}
\BIBentrySTDinterwordspacing

\bibitem{trott2010autodock}
O.~Trott and A.~J. Olson, ``{AutoDock Vina}: Improving the speed and accuracy of docking with a new scoring function, efficient optimization, and multithreading,'' \emph{Journal of Computational Chemistry}, vol.~31, no.~2, pp. 455--461, 2010.

\bibitem{scikit-learn}
F.~Pedregosa, G.~Varoquaux, A.~Gramfort, V.~Michel, B.~Thirion, O.~Grisel, M.~Blondel, P.~Prettenhofer, R.~Weiss, V.~Dubourg, J.~Vanderplas, A.~Passos, D.~Cournapeau, M.~Brucher, M.~Perrot, and E.~Duchesnay, ``Scikit-learn: Machine learning in {P}ython,'' \emph{Journal of Machine Learning Research}, vol.~12, pp. 2825--2830, 2011.

\bibitem{humphrey1996vmd}
W.~Humphrey, A.~Dalke, and K.~Schulten, ``{VMD}: {V}isual {M}olecular {D}ynamics,'' \emph{Journal of Molecular Graphics}, vol.~14, no.~1, pp. 33--38, 1996.

\bibitem{mcmahan2017communication}
B.~McMahan, E.~Moore, D.~Ramage, S.~Hampson, and B.~A.~y. Arcas, ``Communication-efficient learning of deep networks from decentralized data,'' in \emph{Proceedings of the 20th International Conference on Artificial Intelligence and Statistics}.\hskip 1em plus 0.5em minus 0.4em\relax PMLR, 2017, pp. 1273--1282.

\bibitem{pytorch}
A.~Paszke, S.~Gross, F.~Massa, A.~Lerer, J.~Bradbury, G.~Chanan, T.~Killeen, Z.~Lin, N.~Gimelshein, L.~Antiga \emph{et~al.}, ``{PyTorch}: An imperative style, high-performance deep learning library,'' \emph{Advances in Neural Information Processing Systems}, vol.~32, 2019.

\bibitem{ward2021colmena}
\BIBentryALTinterwordspacing
L.~Ward, G.~Sivaraman, J.~G. Pauloski, Y.~Babuji, R.~Chard, N.~Dandu, P.~C. Redfern, R.~S. Assary, K.~Chard, L.~A. Curtiss, R.~Thakur, and I.~Foster, ``Colmena: Scalable machine-learning-based steering of ensemble simulations for high performance computing,'' in \emph{{IEEE}/{ACM} Workshop on Machine Learning in High Performance Computing Environments}.\hskip 1em plus 0.5em minus 0.4em\relax {IEEE}, 2021. [Online]. Available: \url{http://dx.doi.org/10.1109/mlhpc54614.2021.00007}
\BIBentrySTDinterwordspacing

\bibitem{ward2023colmena}
L.~Ward, J.~G. Pauloski, V.~Hayot-Sasson, R.~Chard, Y.~Babuji, G.~Sivaraman, S.~Choudhury, K.~Chard, R.~Thakur, and I.~Foster, ``Cloud services enable efficient {AI}-guided simulation workflows across heterogeneous resources,'' in \emph{Heterogeneity in Computing Workshop}.\hskip 1em plus 0.5em minus 0.4em\relax IEEE Computer Society, 2023, \url{https://arxiv.org/abs/2303.08803}.

\bibitem{berriman2003montage}
G.~Berriman, J.~Good, D.~Curkendall, J.~Jacob, D.~Katz, T.~Prince, and R.~Williams, ``An on-demand image mosaic service for the {NVO},'' in \emph{Astronomical Data Analysis Software and Systems XII}, vol. 295, 2003, p. 343.

\bibitem{montagetutorial}
``{Getting Started: Creating Your First Montage Mosaic},'' \url{http://montage.ipac.caltech.edu/docs/first_mosaic_tutorial.html}. Accessed May 2024.

\bibitem{montagedata}
``{2MASS Image Dataset},'' \url{http://montage.ipac.caltech.edu/docs/Kimages.tar}. Accessed May 2024.

\bibitem{keahey2020lessons}
K.~Keahey, J.~Anderson, Z.~Zhen, P.~Riteau, P.~Ruth, D.~Stanzione, M.~Cevik, J.~Colleran, H.~S. Gunawi, C.~Hammock, J.~Mambretti, A.~Barnes, F.~Halbach, A.~Rocha, and J.~Stubbs, ``Lessons learned from the {C}hameleon testbed,'' in \emph{USENIX Annual Technical Conference}.\hskip 1em plus 0.5em minus 0.4em\relax USENIX Association, July 2020.

\end{thebibliography}
